\documentclass{article}

\usepackage{graphicx}
\usepackage{float}
\usepackage[
singlelinecheck=false 
]{caption}
\usepackage{epsfig}
\begin{document}
\title{\bf Thin accretion disks and  charged rotating dilaton black holes}
\author{{ Mohaddese Heydari-Fard$^{1}$ \thanks{Electronic address: m\_heydarifard@sbu.ac.ir}, Malihe Heydari-Fard$^{2}$\thanks{Electronic address:heydarifard@qom.ac.ir } and Hamid Reza Sepangi$^{1}$\thanks{Electronic address: hr-sepangi@sbu.ac.ir}}\\{\small \emph{$^{1}$ Department of Physics, Shahid Beheshti University, G. C., Evin, Tehran 19839, Iran}}\\{\small \emph{$^{2}$ Department of Physics, The University of Qom, 3716146611, Qom, Iran}}}
\maketitle

\begin{abstract}
Einstein-Maxwell-dilaton theory is an interesting theory of gravity for studying scalar fields in the context of no-hair theorem. In this work, we consider static charged dilaton and charged, slowly rotating dilaton black holes in Einstein-Maxwell-dilaton gravity. We investigate the accretion process in thin disks around such black holes, using the Novikov-Thorne model. The electromagnetic flux, temperature distribution, energy conversion efficiency and also innermost stable circular orbits of thin disks are obtained and effects of dilaton and rotation parameters are studied. For the static and slowly rotating black holes the results are compared to that of Schwarzschild and Kerr, respectively.
\vspace{5mm}
\vspace{1mm}
\end{abstract}

\section{Introduction}
General relativity (GR) is a successful theory for describing  various gravitational phenomena from  planetary to cosmic scales. In contrast, at very small scales where quantum gravity becomes essential, there is no such viable theory. Nevertheless, string theory somewhat paves the way for quantum gravity by supplementing the usual Einstein-Hilbert action with possible higher-order curvature invariants which result from its low energy limit, together with an additional scalar dilaton field non-minimally coupled to gravity \cite{1}. These considerations have motivated a large number of studies in the past decades in a scenario where a dilaton field is non-minimally coupled to a Maxwell field, also known as Einstein-Maxwell-dilaton gravity (EMDG).

In the absence of dilaton potential, charged dilaton black holes and slowly rotating charged dilaton black holes, as exact solutions of EMDG, have been studied in \cite{2}, \cite{3} and \cite{d22}, respectively. All these  solutions are asymptotically flat. Also, recently the authors in \cite{ref1}--\cite{ref2} have obtained exact asymptotically flat hairy black holes with a dilaton potential. These solutions can recover some solutions of \cite{2} and Reissner-Nordstrom solution in the limiting cases of a dilaton potential. They can be overcharged and cure the inner horizon problem of dilaton black holes and are thermodynamically and dynamically stable \cite{ref3}-\cite{ref4}. The effect of the hair of such asymptotically flat black holes on the bending angle of light have been considered in \cite{light}. On the other hand, in \cite{A19}--\cite{A20}, by considering a suitable combination of three Liouville-type potentials, exact charged dilaton black hole solutions in an (anti)-de Sitter space-time have been constructed in the presence of linear Maxwell electrodynamics. Generalization of this study for the case of nonlinear electrodynamics has been investigated in \cite{sheykhi21}. Also,  black hole solutions  which are neither asymptotically flat nor (A)dS are studied in \cite{13}--\cite{18}. In higher dimensions, slowly rotating charged black hole solutions of EMDG have been studied in \cite{sheykhi22}. Superradiance, fate and evaporation of dilaton black holes have been discussed in \cite{super} and \cite{Maeda}, respectively. Geodesic analyses for the static, rotating, electrically and magnetically charged dilaton black holes have been carried out in \cite{d23}--\cite{d25}. Also, light path and deflection angle in normal and phantom dilaton black holes have been studied in \cite{d26}. Phase transition and thermodynamical behavior of charged dilaton black holes have been considered in \cite{d27}. New solutions of dilaton black holes in the presence of Gauss-Bonnet term were done in \cite{d28}. Quasinormal modes and shadows cast by dilaton black holes and dilaton wormholes are studied in \cite{d29}--\cite{d30}.

A black hole has no electromagnetic radiation and can only be detected through its gravitational effects on the surrounding objects. We can observe the stellar-mass black holes in some special cases. For instance, the discovery of gravitational waves resulting from merging of a binary black hole system provided the best evidence for the existence of black holes \cite{b1}--\cite{b2}. Also, a stellar-mass black hole can be found through the X-ray emission of a binary system. In such binaries the other component is a visible star from which there is an outflow of matter to the black hole. This process leads to formation of an accretion disk around the black hole and one may observe the X-ray radiation from the inner part of the disk \cite{x1}--\cite{x2}. In non-interacting binaries the presence of a black hole can be inferred by the radial velocity measurements of the companion star \cite{v1}--\cite{v2}. The first study of accretion disks based on Newtonian approach was carried out in \cite{31}. Subsequently, in several papers by Novikov, Thorne and Page \cite{32}--\cite{34}, the general relativistic approach was investigated. The model is based on the assumption that the mass accretion rate is constant and is not dependent on the radius of the disk, that is, the disk is in a steady state. The accreting matter is  also supposed to move in Keplerian orbit. This assumption requires that the central object does not have a strong magnetic field. Moreover the radiation emanating from the disk  is considered under the assumption of black body radiation resulting from thermodynamic equilibrium of the disk. Thin accretion disk models and their properties in $f(R)$ modified gravity have been investigated in \cite{35}--\cite{37}.  In extra dimensions within the context of modified gravity, such as Kaluza-Klein and brane world scenarios, accretion disks have been studied in \cite{38}--\cite{40}. Accretion disks in Chern-Simons and scalar-tensor-vector gravity have been considered in \cite{chern}--\cite{41}. The analysis of disk properties around an exotic matter such as wormholes is an interesting subject that has been considered in \cite{42}--\cite{new-wormhole}. Accretion disks for other compact astrophysical objects such as neutron, boson and fermion stars and gravastars  have  been studied in \cite{44}--\cite{52}. In a recent work \cite{new}, the authors have studied thin accretion disks around electrically and magnetically charged Gibbons-Maeda-Garfinkle-Horowitz-Strominger (GMGHS) black holes. Also, $k_{\alpha}$ iron line analysis and continuum-fitting method are used to distinguish different astrophysical objects through their accretion disks \cite{53}--\cite{56}. In this paper we consider static and slowly rotating charged dilaton black holes and study the properties of thin accretion disks around them.

The structure of the paper is as follows. In section 2, we give a brief review of geodesic equations and accretion process in thin disks around a general stationary axisymmetric space-time. In section 3 we introduce static and charged rotating dilaton black holes and  derive the effective potential,  electromagnetic flux, temperature distribution and energy conversion efficiency of thin disks in the context of EMDG and move on to investigate the effects of dilaton coupling $\alpha$ and rotation parameter $a$ on the disk properties. The concluding remarks are presented in the last section.

\section{Accretion in thin disks around stationary axisymmetric spacetimes}
Let us first briefly review the electromagnetic radiation properties of a thin accretion disk in  general stationary axisymmetric space-times \cite{chern} and derive the basic mathematical equations that are needed to study Einstein-Maxwell-dilaton black hole solutions.

\subsection{Geometry of space-time and geodesic equations}
Accretion disks form by particles moving on  geodesics orbiting  a compact central object. The general form of the metric for a stationary axisymmetric space-time is given by
\begin{equation}
ds^2=g_{tt}dt^2+2g_{t\phi}dtd\phi+g_{rr}dr^2+g_{\theta\theta}d\theta^2+g_{\phi\phi}d\phi^2.
\label{1}
\end{equation}
In equatorial approximation, $\mid\theta-\frac{\pi}{2}\mid \ll 1$, we assume that $g_{tt}, g_{rr}, g_{\theta\theta}$, $g_{\phi\phi}$ and $g_{t\phi}$ components of the metric are functions of $r$ only. The geodesic equations in terms of the constants of motion, namely the specific energy ${\tilde{E}}$ and the specific angular momentum ${\tilde{L}}$, using the above metric are given by
\begin{equation}
\frac{dt}{d\tau}=\frac{\tilde{E}g_{\phi\phi}+\tilde{L}g_{t\phi}}{g_{t\phi}^2-g_{tt}g_{\phi\phi}},
\label{2}
\end{equation}
\begin{equation}
\frac{d\phi}{d\tau}=-\frac{\tilde{E}g_{t\phi}+\tilde{L}g_{tt}}{g_{t\phi}^2-g_{tt}g_{\phi\phi}},
\label{3}
\end{equation}
and
\begin{equation}
g_{rr}\left(\frac{dr}{d\tau}\right)^2=V_{eff}(r),
\label{4}
\end{equation}
where $\tau$ is the affine parameter. The normalization condition, $g_{\mu\nu}u^{\mu}u^{\nu}=-1$, for test particles leads to the following effective potential
\begin{equation}
V_{eff}(r)=-1+\frac{\tilde{E}^2g_{\phi\phi}+2\tilde{E}\tilde{L}g_{t\phi}+
\tilde{L}^2g_{tt}}{g_{t\phi}^2-g_{tt}g_{\phi\phi}}.
\label{5}
\end{equation}
For circular orbits with an arbitrary radius $r$ in equatorial plane we have $V_{eff}(r)=0$ and $V_{eff,r}(r)=0$. Use of these conditions leads to the angular velocity $\Omega$, the specific angular momentum ${\tilde{L}}$ and the specific energy ${\tilde{E}}$ for a particle in a circular orbit in the gravitational potential of a massive object
\begin{equation}
{\tilde{E}}=-\frac{g_{tt}+g_{t\phi}\Omega}{\sqrt{-g_{tt}-2g_{t\phi}\Omega-g_{\phi\phi}\Omega^2}},
\label{6}
\end{equation}
\begin{equation}
{\tilde{L}}=\frac{g_{t\phi}+g_{\phi\phi}\Omega}{\sqrt{-g_{tt}-2g_{t\phi}\Omega-g_{\phi\phi}\Omega^2}},
\label{7}
\end{equation}
\begin{equation}
\Omega=\frac{d\phi}{dt}=\frac{-g_{t\phi,r}+\sqrt{(g_{t\phi,r})^2-g_{tt,r}g_{\phi\phi,r}}}{g_{\phi\phi,r}}.
\label{8}
\end{equation}
Moreover, to determine the inner edge of the disk we should determine the innermost stable circular orbit (ISCO) of the black hole potential by using the condition $V_{eff,rr}(r)=0$ which leads to the following relation
\begin{equation}
{\tilde{E}^2g_{\phi\phi,rr}}+2\tilde{E}\tilde{L}g_{t\phi,rr}+{\tilde{L}^2g_{tt,rr}}-\left(g_{t\phi}^2-
g_{tt}g_{\phi\phi}\right)_{,rr}=0. \label{9}
\end{equation}
As can be seen, the above quantities depend only on the metric components.  In the next section we determine these quantities and the ISCO radius for Einstein-Maxwell-dilaton black holes.

\subsection{Electromagnetic properties and structure of thin accretion disks}
Let us now summarize the physical properties of thin accretion disks which we will use in our calculations, such as the energy flux emitted by the disk, $F(r)$, the efficiency $\epsilon$, temperature distribution $T(r)$ and Luminosity spectra $L(\nu)$. We use the standard relativistic thin accretion disk model developed by Novikov and Thorne \cite{32} which is a generalization of that studied by Shakura-Sunyaev \cite{31}. In what follows we state the assumptions used in Novikov-Thorne model.

The first assumption is that the accretion disk is geometrically thin and optically thick. This respectively means that the vertical size of the  disk, $h$, is negligible compared to its horizontal size, $h\ll r$, and that the photon mean free path in the disk, $l$, is negligible  compared to its depth, $l\ll h$. The second assumption is that the space-time is stationary, axisymmetric, asymptotically flat and reflection-symmetric in the equatorial plane and that the self-gravity of the disk is negligible. The third assumption states that the disk is in both hydrodynamical and thermodynamic equilibrium; the electromagnetic properties of the disk is similar to a black body and that the disk's temperature does not increase by converting all the gravitational energy into heat.
The fourth assumption is that the disk lies in the equatorial plane of the accreting compact object and has an inner edge in a marginally stable orbit, $r_{ms}$, known as the ISCO radius and extends to the outer edge, $r_{out}$. Disk particles have Keplerian motion between $r_{isco}$ and $r_{out}$ accreted by the central massive object. Finally, the last assumption considers disks  to be in a steady-state. In the steady-state of Novikov-Thorne model it is assumed that the mass accretion rate, $\dot{M}_0$, is constant and does not change with time. In the standard thin accretion disk the physical quantities which describe the orbiting plasma are averaged over the time scale $\triangle t$ for a total period of the orbits and over the thickness of the disk $H$.

The radiation flux emitted by the surface of accretion disk can be derived from three structure equations, namely the conservation equation of rest mass, energy and the angular momentum of disk particles \cite{book}
\begin{equation}
F(r)=\frac{\dot{M}_{0}}{4\pi M^2} f_{disk}(r),
\label{10}
\end{equation}
where
\begin{equation}
 f_{disk}(r)\equiv-\Omega_{,r}\frac{M^2}{\sqrt{-g}\left(\tilde{E}-\Omega \tilde{L}\right)^2}\int^r_{r_{isco}}\left(\tilde{E}-\Omega \tilde{L}\right) \tilde{L}_{,r}dr.
\label{11}
\end{equation}
Here, $\dot{M}_{0}$ is the mass accretion rate and $f_{disk}(r)$ is a dimensionless function of radius.

As was mentioned above, the thin disk is assumed to be in local thermal equilibrium and so the radiation emitted by the disk surface is black body-like for which the Stefan-Bltzmann law is valid
\begin{equation}
F(r)=\sigma_{SB} T^4(r),
\label{12}
\end{equation}
where $\sigma_{SB}=5.67\times10^{-5} ergs^{-1} cm^{-2} k^{-4}$ is the Stefan-Boltzmann constant. The observed luminosity $L(\nu)$ has a red-shifted black body spectrum
\begin{equation}
L(\upsilon)=4\pi d^2 I(\nu)=\frac{8\pi h \cos\gamma}{c^2}\int_{r_{in}}^{r_{out}}\int_0^{2\pi}\frac{ \nu_e^3 r dr d\phi}{\exp{[\frac{h\nu_e}{k_{B} T}]}-1},\label{13}
\end{equation}
where $h$ is the Plank constant, $k_{B}$ is the Boltzmann constant, $\gamma$ is the disk inclination angle and $r_{in}$ and $r_{out}$ are inner and outer radii of the edge of disk.

Radiative efficiency is another parameter in accreting process which demonstrates the conversion of rest mass into radiation by the central object. When the emission of disk and the absorption by the black hole are negligible, the Novikov-Thorne radiative efficiency is given by
\begin{equation}
\epsilon=1-\tilde{E}_{isco}.
\label{14}
\end{equation}
It is important to point out that the accretion mechanism is an efficient process for converting rest mass into radiation. Although in the case of a star the efficiency is less than 1\% due to nuclear fusion, in a Schwarzschild black hole about 6\% of rest mass is converted into energy. Also in a rotating black hole, without capture of radiation by the hole, the efficiency can be 42\% and, if the photon capture is taken into account, it is about 40\% if the spin parameter of the black hole is $a=1$.

\section{Properties of thin accretion disks in EMD gravity}
\subsection{Static charged dilaton black holes}
The action of Einstein-Maxwell-dilaton gravity is given by
\begin{equation}
{\cal S}=\int d^4x \sqrt{-g}\left[R-2g^{\mu \nu}\nabla_{\mu}\Phi\nabla_{\nu}\Phi-e^{-2\alpha\Phi}F_{\mu \nu}F^{\mu \nu}\right],
\label{15}
\end{equation}
where the Ricci scalar and dilaton field are represented by $R$ and $\Phi$ respectively and $F_{\mu\nu}$ is the usual electromagnetic tensor  defined as $F_{\mu\nu}=\partial_{\mu}A_{\nu}-\partial_{\nu}A_{\mu}$ with $A_{\mu}$ being the electromagnetic vector potential. The dilaton coupling constant $\alpha$ represents the strength with which the dilaton is coupled to the Maxwell field. For $\alpha=0$ this is the effective action of Einstein-Maxwell theory coupled to a dilaton scalar field. In this case, according to no-hair conjecture, $\Phi$ must be constant and static black hole solutions are the same as that of the Reissner-Nordstrom. The low energy limit of the string theory can be achieved for $\alpha=1$ and the case of $\alpha=\sqrt{3}$ corresponds to the five-dimensional Kaluza-Klein theory.

Static, spherically symmetric charged black hole solutions governed by action (\ref{15}) were presented in \cite{2}. The line element is
\begin{equation}
ds^2=-f(r)dt^2+\frac{dr^2}{f(r)}+R^2(r)\left(d\theta^2+\sin ^2\theta d\phi^2\right),
\label{16}
\end{equation}
where
\begin{equation}
f(r)=\left(1-\frac{r_+}{r}\right)\left(1-\frac{r_{-}}{r}\right)^{\frac{1-\alpha^2}{1+\alpha^2}},
\label{17}
\end{equation}
and
\begin{equation}
R^2(r)=r^2\left(1-\frac{r_{-}}{r}\right)^{\frac{2\alpha^2}{1+\alpha^2}}.
\label{18}
\end{equation}
We have denoted the radius of the outer and inner event horizons by $r_{+}$ and $r_{-}$, given by
\begin{equation}
r_{+}= M[1+\sqrt{1-v^2(1-\alpha^2)}],
\label{19}
\end{equation}
\begin{equation}
r_{-}=\frac{ M(1+\alpha^2)[1-\sqrt{[1-v^2(1-\alpha^2)}]}{(1-\alpha^2)},
\label{20}
\end{equation}
where $M$ and $v=\frac{Q}{M}$ are the black hole mass and electric charge to mass ratio, respectively. Moreover, the maximal value of charge, i.e. $Q_{max}=M\sqrt{1+\alpha^2}$ can be obtained when $r_+=r_-$ and for this reason the  black hole is known as extremal. The extremal limit of the static solution (\ref{16}) is not well defined because the event horizon $r_+=M(1+\alpha^2)$ becomes a naked singularity and it has vanishing area for $\alpha\neq 0$. Also in this limit there is a critical value of the coupling constant at $\alpha=1$, so that when $\alpha<1$ the temperature in the extreme limit vanishes and for $\alpha >1$ it diverges. However for $\alpha=1$ it remains finite  \cite{Maeda}.

As is clear, in the limiting case $v=0$ with arbitrary values of $\alpha$, the above metric reduces to the Schwarzschild metric where $r_{+}=2M$ and $r_{-}=0$ are the coordinate and intrinsic singularity respectively. Also, for $\alpha=0$ it reduces to Reissner-Nordstrom metric with $r_{\pm}=M[1\pm\sqrt{1-v^2}]$.
The behaviour of the dilaton field and vector potential are given by
\begin{equation}
\Phi(r) = \frac{\alpha}{1+\alpha^2}\log{\left(1-\frac{r_{-}}{r}\right)},
\label{21}
\end{equation}
\begin{equation}
A_{t} = \frac{Mv}{r}.
\label{22}
\end{equation}
The dilaton charge defined over a two-sphere at infinity is \cite{2}
\begin{equation}
D(r)=-r^2 \Phi(r)_{,r}\mid_{r\rightarrow\infty}.
\label{23}
\end{equation}
Now, using the metric components of equation (\ref{16}), we obtain the specific energy, specific angular momentum and angular velocity of EMD black holes as given below
\begin{equation}
{\tilde{E}}=\frac{\sqrt{2}(1-\frac{r_+}{r})(1-\frac{r_-}{r})^{\frac{1-\alpha^2}{1+\alpha^2}}}{\sqrt{\frac{
(1-\frac{r_-}{r})^{\frac{1-\alpha^2}{1+\alpha^2}}[2(1+\alpha^2)r^2+r(\alpha^2(r_--3r_+)-3(r_++r_-))+4r_+r_-]}
{r(\alpha^2r+r-r_-)}}},
\label{24}
\end{equation}
\begin{equation}
{\tilde{L}}=\frac{r^2(1-\frac{r_-}{r})^{\frac{2\alpha^2}{1+\alpha^2}}{\sqrt{\frac{
(1-\frac{r_-}{r})^{\frac{1-3\alpha^2}{1+\alpha^2}}[r(\alpha^2(r_+-r_-)+r_++r_-)-2r_+r_-]}
{r^3(\alpha^2r+r-r_-)}}}}{\sqrt{\frac{
(1-\frac{r_-}{r})^{\frac{1-\alpha^2}{1+\alpha^2}}[2(1+\alpha^2)r^2+r(\alpha^2(r_--3r_+)-3(r_++r_-))+4r_+r_-]}
{r(\alpha^2r+r-r_-)}}},\label{25}
\end{equation}
and
\begin{equation}
\Omega={\sqrt{\frac{\left(1-\frac{r_-}{r}\right)^{\frac{1-3\alpha ^2}{1+\alpha ^2}} \left(r \left(\alpha ^2 ({r_+}-{r_-})+{r_+}+{r_-}\right)-2 {r_+} {r_-}\right)}{2r^3 \left(\alpha ^2 r+r-{r_-}\right)}}}.
\label{26}
\end{equation}
The ISCO equation (\ref{9}) is now given by
\begin{equation}
c_{1}r^{4}+c_{2}r^{3}+c_{3}r^{2}+c_{4}r+c_{5}=0,
\label{isco}
\end{equation}
where
\begin{eqnarray}
c_1 &=&-2(1+\alpha^2)^2\left[(r_{+}+r_-)+\alpha^2(r_+-r_-)\right],\nonumber\\
c_2 &=&2(1+\alpha^2)\left[(r_++r_-)+\alpha^2(r_+-r_-)\right]\left[3r_+(1+\alpha^2)+2r_-(2-\alpha^2)\right], \nonumber\\
c_3 &=&-2r_{-}\left[12r_{+}^{2}(1+\alpha^2)^{2}+15r_{+}r_{-}(1-\alpha^4)+r_{-}^{2}(3-4\alpha^2+\alpha^4)\right], \nonumber\\
c_4 &=&2r_{+}r_{-}^2\left[r_{+}(1+\alpha^2)(17-\alpha^2)+r_{-}(9-8\alpha^2-\alpha^4))\right], \nonumber\\
c_5 &=&-16r_{+}^{2}r_{-}^{3},\label{i2}
\end{eqnarray}
and $r_+$ and $r_-$ are given by equations (\ref{19}) and (\ref{20}). Only one of the real roots of this equation is outside the event horizon and determines the ISCO radius of the accretion disk. For small $\alpha$, the above equations can be used for an expansion to ${\cal O} (\alpha^2)$ to find the corrections to radius of the accretion disk and comparison with that of black holes when $\alpha$ vanishes. Meanwhile it is easy to verify that in the case $v=0$ it reduces to ISCO equation for the Schwarzschild black hole with $r_{isco}=6M$.

We have calculated  these quantities for a EMD black hole for dilaton coupling $\alpha=0.2$ and different values of $v$. The results are displayed in figure 1 and for comparison we have also plotted the corresponding result for a disk rotating around a Schwarzschild black hole in GR. We see that both $\tilde{E}$ and $\tilde{L}$ are larger in GR than in EMDG and, as the value of parameter $v$ increases, deviation with respect to GR becomes more prominent, although the angular velocity in EMDG is almost the same as in GR. Also the radial profile of the effective potential is given by
\begin{equation}
V_{eff}(r)=\left(1-\frac{r_+}{r}\right)\left(1-\frac{r_-}{r}\right)^{\frac{1-\alpha^2}{1+\alpha^2}}\left(1+
\frac{\tilde{L^2}}{r^2}(1-\frac{r_-}{r})^{\frac{-2\alpha^2}{1+\alpha^2}}\right).
\label{27}
\end{equation}
The behavior of the effective potential for $\alpha=0.2$ and different values of $v$ for a EMD black hole is plotted in figure 1. For smaller values of $v$ the peak of the effective potential decrease so that the Schwarzschild black hole has a minimum peak compared to EMD black holes.

\begin{figure}[H]
\centering
\includegraphics[width=3.0in]{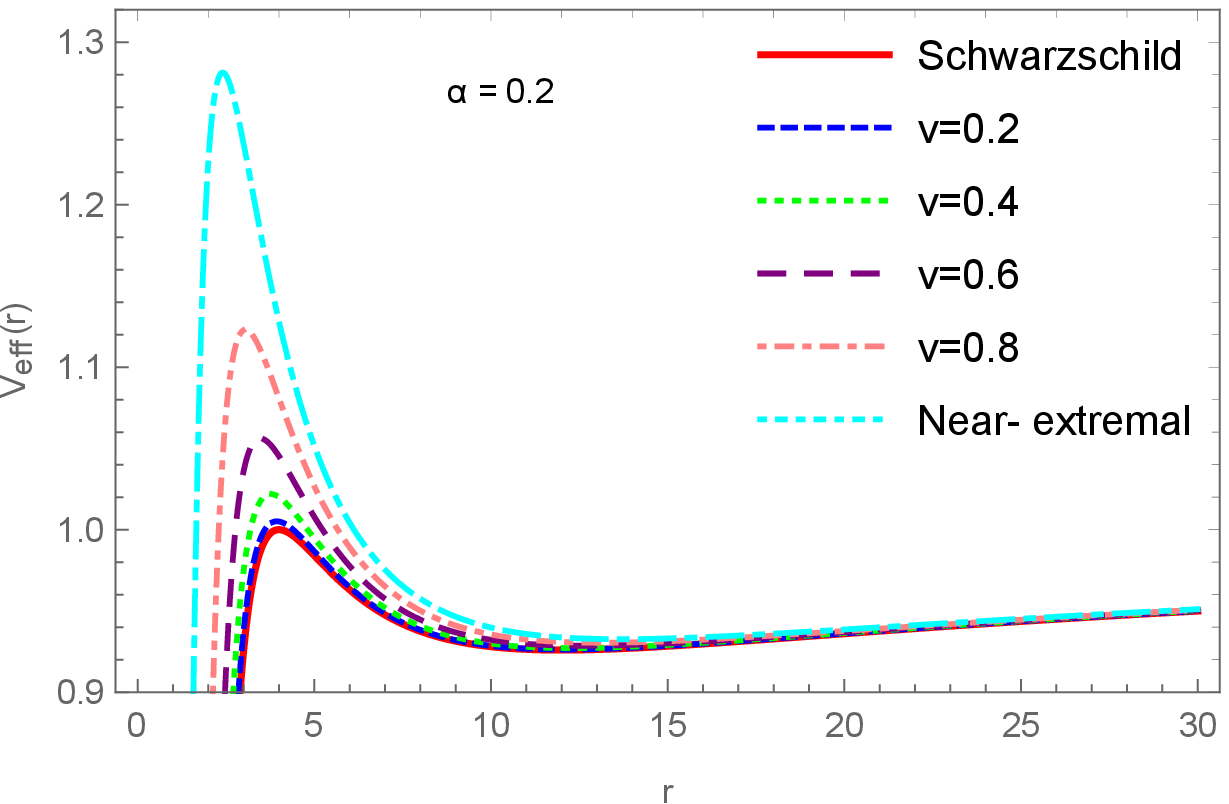}
\includegraphics[width=3.0in]{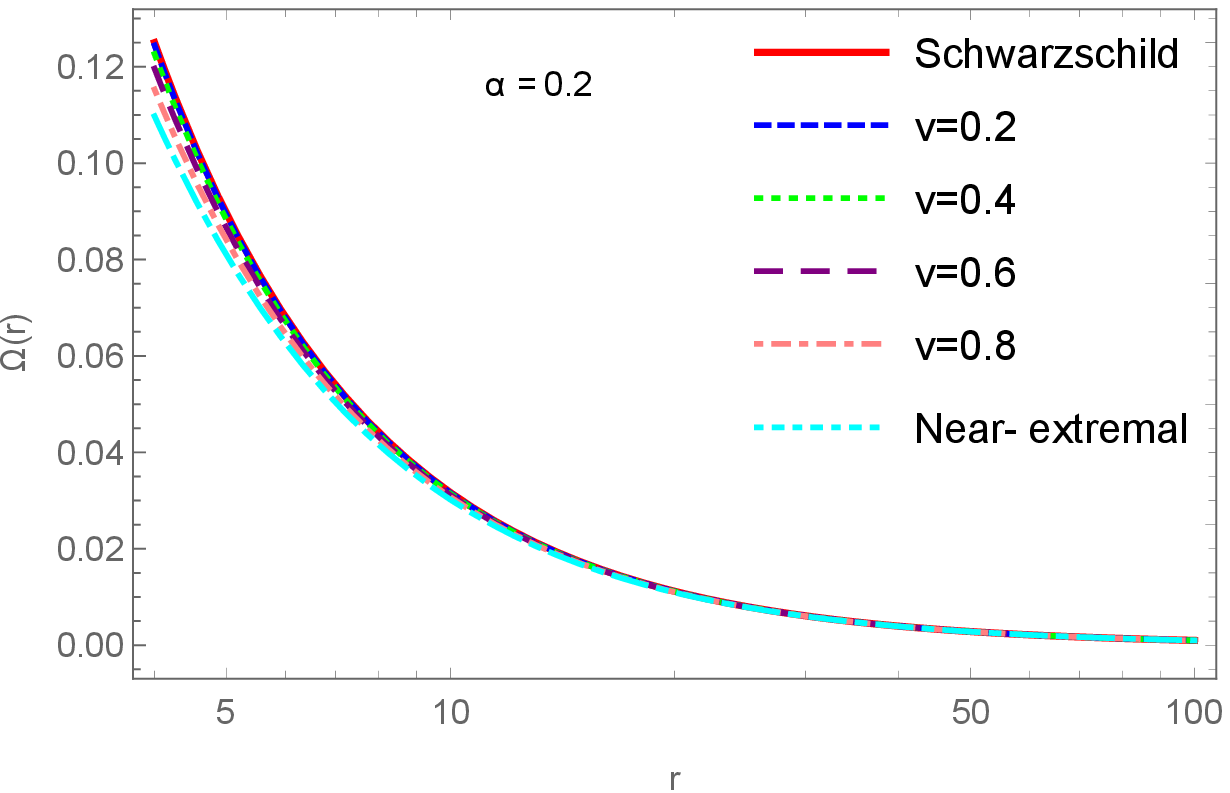}\\
\includegraphics[width=3.0in]{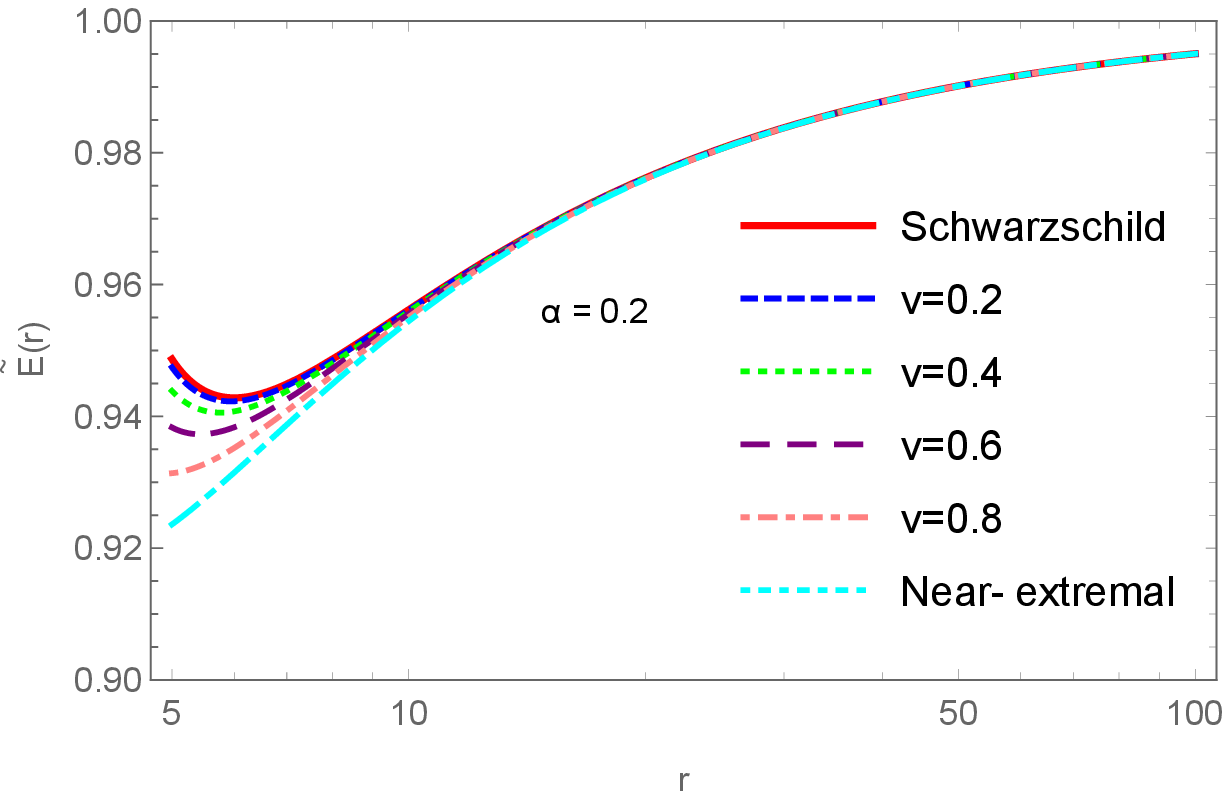}
\includegraphics[width=3.0in]{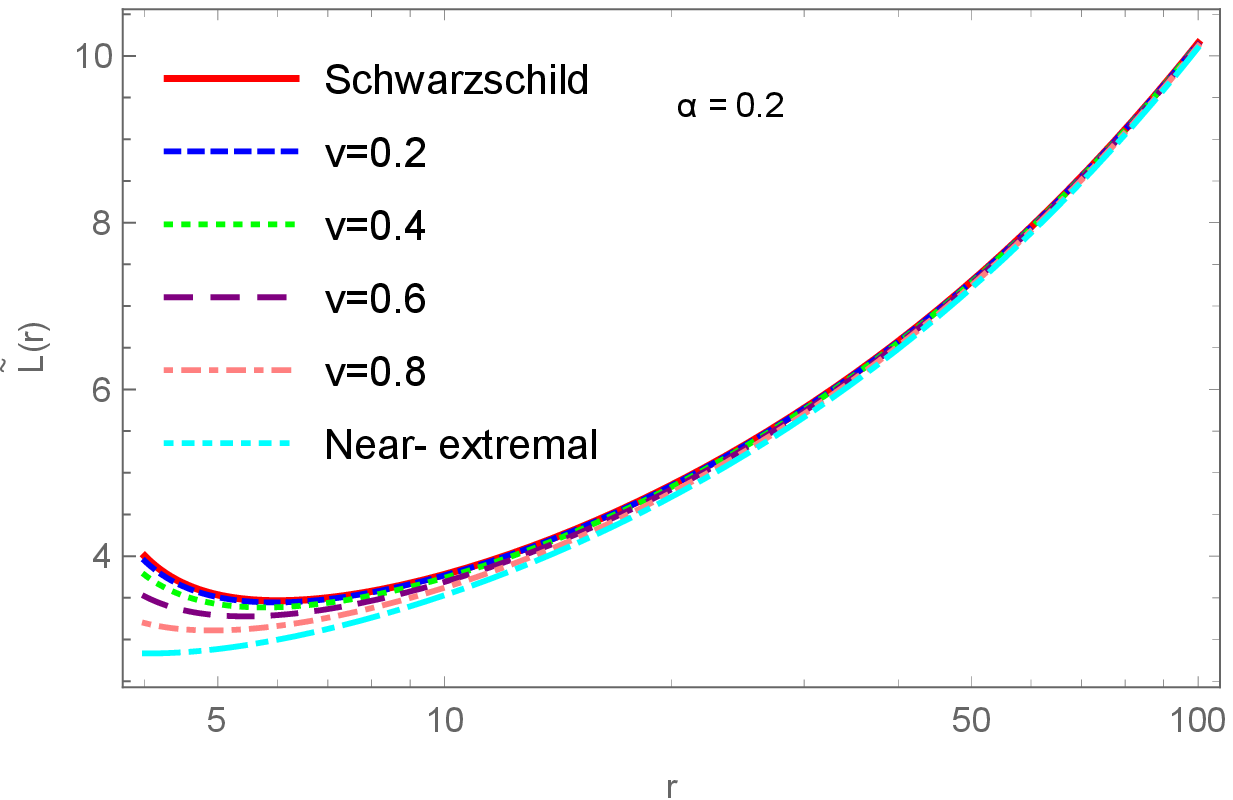}
\caption{The effective potential $V_{eff}(r)$. Top-left panel: the angular velocity $\Omega(r)$. Top-right panel: the specific energy $\tilde{E}(r)$. Bottom-left panel: the specific angular momentum $\tilde{L}(r)$. Bottom-right panel:  a static charged dilaton black hole with total mass $M=2.5\times10^6M_{\odot}$  shown as a function of the radial coordinate $r$ for different values of $v$ and are compared with the Schwarzschild black hole ($v=0$). At the near-extremal limit, $Q_{max}\sim M\sqrt{1+\alpha^2}$, the effective potential has the largest maximum. The dilaton parameter is set to $\alpha=0.2$.}
\label{potential}
\end{figure}

The radial profile of the energy flux over the surface of the disk is shown in figure 2. In the left panel, we have displayed this quantity for a fixed value of $\alpha$ and different values of $v$. As is clear, the energy flux is larger in EMDG than in GR and, as the value of  parameter $v$ increases  deviation from GR also increases,  similar to the effective potential. In addition, the figure shows that close to extremity, EMD black holes have the largest maximum of energy flux. In order to see the effect of dilaton parameter on the energy flux in the right panel of figure 2 we have plotted the energy flux for a fixed vale of $v$ and different values of $\alpha$ . The dilaton parameter runs from $\alpha=0$, representing a Reissner-Nordstrom solution, to $\alpha=2$. It should be noted that for $\alpha=1$, the low energy limit of the string theory, we have used the GMGHS metric where the disk properties of this special case have been studied in \cite{new}. The disk temperature profile is displayed in figure 3 and the same features are also observed.

In table 1, we present the ISCO radius, $r_{isco}$ and efficiency of the EMD black hole, $\epsilon$, for different values of $\alpha$ and $v$. We find the ISCO radius by calculating the roots of equation (\ref{9}) where only one of the real roots is outside the event horizon of the EMD black kole. According to table 1 for a fixed value of $\alpha$ the ISCO radius increase as the value of $v$ decreases so that for the Schwarzschild space-time with smallest value of $v$ ($v=0$), the ISCO radius has the largest value $r_{isco}=6 M$ and as we expect it has the lowest value of the energy flux, temperature and efficiency, as shown in the left panel of figure 2. This is because the gravitational field in GR is stronger than that in  EMDG so that the instability area around the black hole increases and the ISCO radius assumes larger values.

\begin{figure}[H]
\centering
\includegraphics[width=3.0in]{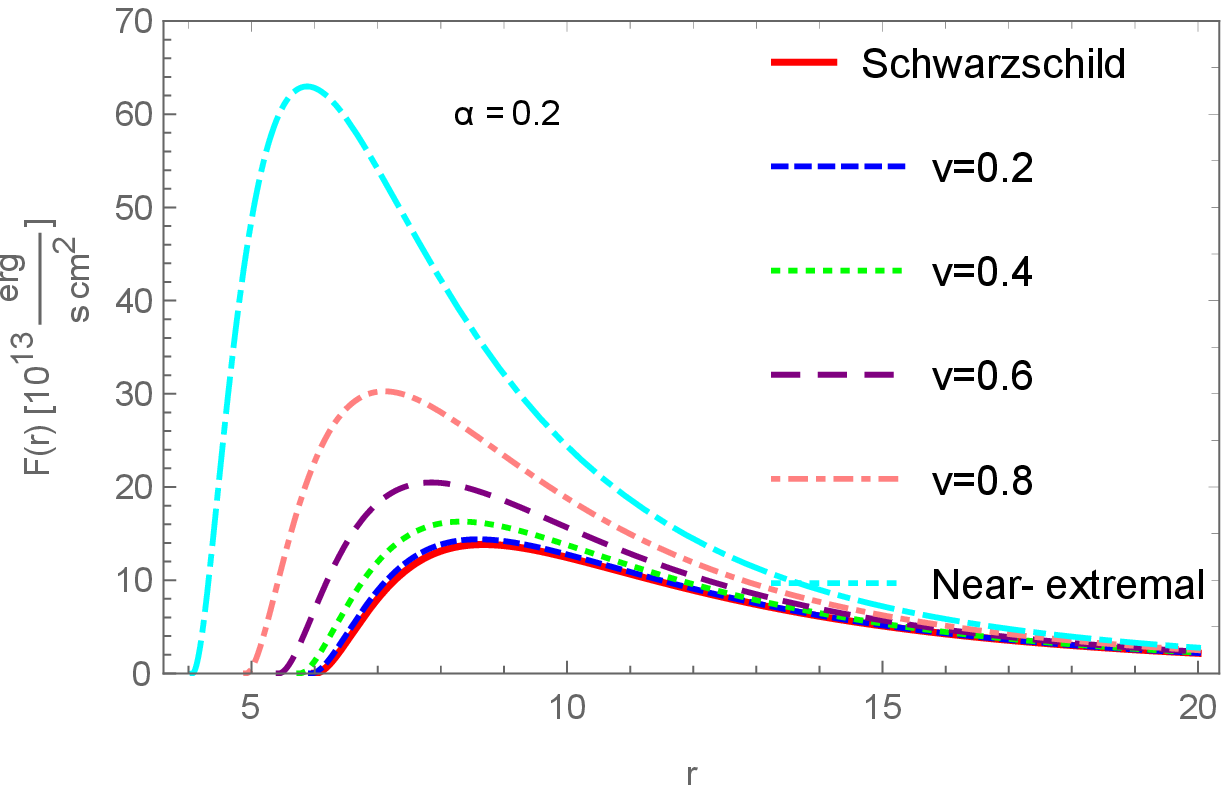}
\includegraphics[width=3.0in]{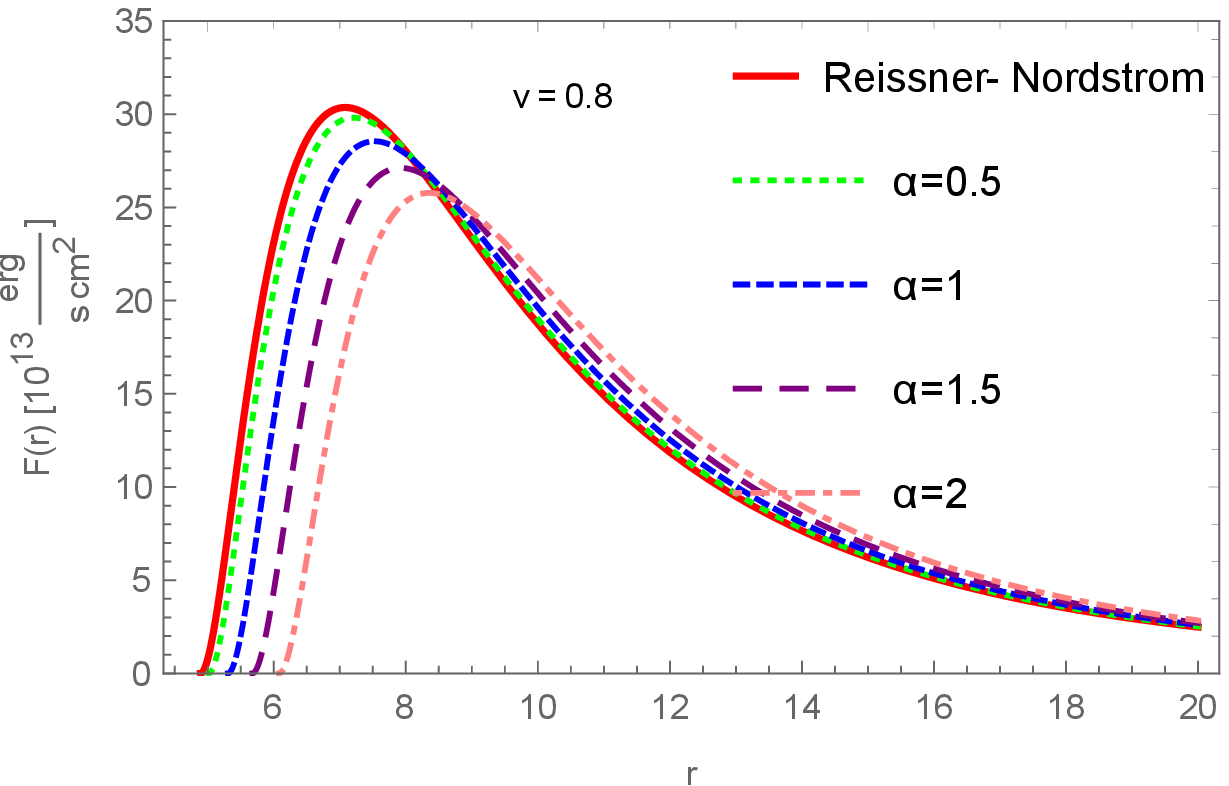}
\caption{The energy flux $F(r)$ of a disk around a static charged dilaton black hole with the mass accretion rate $\dot{M}=2\times10^{-6}M_{\odot}yr^{-1}$, for different values of $v$, left panel and different values of $\alpha$, right panel. The solid curves correspond to the Schwarzschild ($v=0$) and Reissner-Nordstrom ($\alpha=0$) black holes in the left and right panels, respectively. At the near-extremal limit $Q_{max}\sim M\sqrt{1+\alpha^2}$,  the energy flux has the largest maximum.}
\label{flux}
\end{figure}
\begin{figure}[H]
\centering
\includegraphics[width=3.0in]{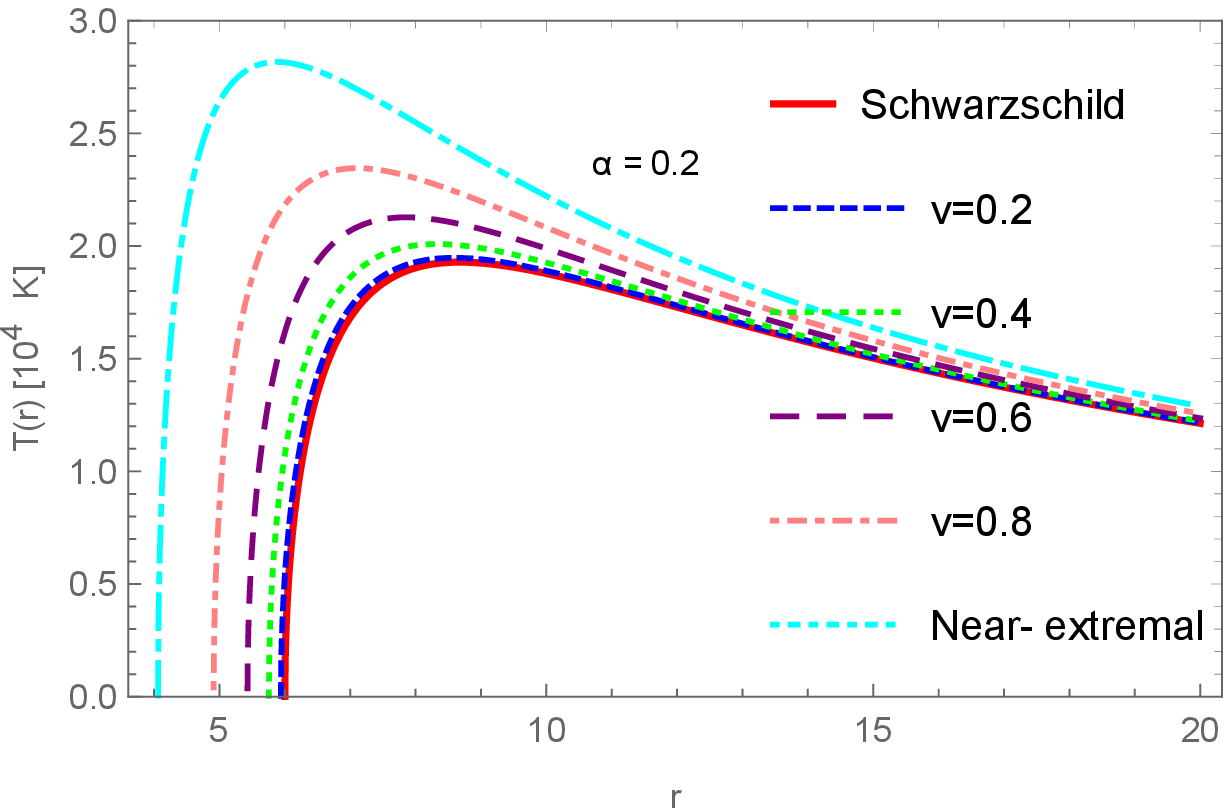}
\includegraphics[width=3.0in]{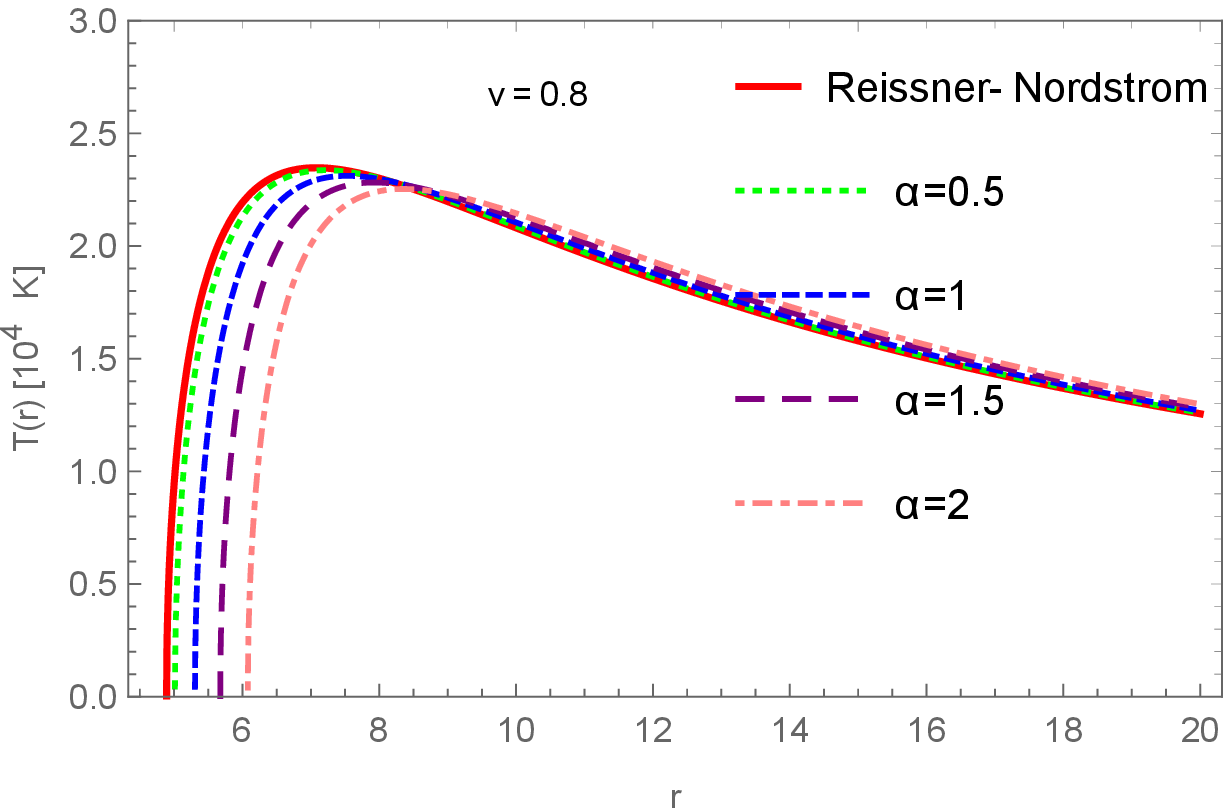}
\caption{Disk temperature $T(r)$ for a static charged dilaton black hole with mass accretion rate $\dot{M}=2\times10^{-6}M_{\odot}yr^{-1}$, for different values of $v$, left panel and different values of $\alpha$, right panel. The solid curves correspond to Schwarzschild ($v=0$) and Reissner-Nordstrom ($\alpha=0$) black holes in the left and right panels, respectively. At the near-extremal limit $Q_{max}\sim M\sqrt{1+\alpha^2}$, the disk temperature has the largest maximum.}
\label{Temperature}
\end{figure}

Table 1 shows that for a fixed value of $v$, as the value of $\alpha$ decrease the value of ISCO radius also decreases and tends to the limiting case of Reissner-Nordstrom ($\alpha=0$) with $r_{isco}=4M$. Therefore the smaller the dilaton parameter, the larger the efficiency. We note that the rate of increase in efficiency is smaller for the smallest values of $v$.

In the extremal limit the ISCO radius, $r_{isco}$, and efficiency of an EMD black hole are presented in table 3. The results show that in the extreme case the inner edge of the disk approaches the event horizon and the black hole becomes more efficient as shown in the left panels of figure 2 and 3.

\begin{table}[H]
\centering
\caption{The $r_{isco}$ of the accretion disk and efficiency for a static, charged dilaton black hole.}
\begin{tabular}{|l|l|l|l|l|l|}
\hline
$\alpha$&$v$&$r_{isco}/M$& $r_{+}/M$&$\epsilon$\\ [0.5ex]
\hline
{0.2}  &0.2& 5.9404&1.9806& 0.0577\\
       &0.4& 5.7563&1.9200& 0.0594\\
       & 0.6& 5.4289&1.8089&0.0627\\
        &0.8& 4.9115&1.6209& 0.0687\\
         &0.9& 4.5465&1.4715& 0.0735\\
\hline
{0.4}
     &0.2& 5.9428&1.9830& 0.0577\\
     &0.4& 5.7668&1.9303& 0.0594\\
     &0.6& 5.4556&1.8352& 0.0627\\
     &0.8& 4.9708&1.6800& 0.0685\\
     &0.9& 4.6364&1.5653& 0.0732\\
\hline
{0.6}
&0.2& 5.9469&1.9871& 0.0577\\
&0.4& 5.7839&1.9474&0.0594\\
&0.6& 5.4984&1.8773&0.0626\\
&0.8& 5.0612&1.7684&0.0683\\
&0.9& 4.7669&1.6939& 0.0728\\
\hline
{1.2}
&0.2& 5.9686&2.0088&0.0577\\
&0.4& 5.8720&2.0346&0.0594\\
&0.6& 5.7028&2.0763&0.0624\\
&0.8& 5.4475&2.1321&0.0674\\
&0.9& 5.2807&2.1646&0.0710\\
\hline
{-}
&0& 6&2&0.0572\\
\hline
\end{tabular}
\end{table}

The EMD black holes we are studying have a non-trivial profile of scalar field outside the event horizon. Therefore, they have a scalar hair, as defined in equation (\ref{23}). This dilaton hair, however, is not an independent quantity and as we have shown in figure 4,  depends on the black hole mass. Hence, the dilaton hair of such black holes is a ``secondary hair'' associated with the primary hair (mass). As is clear from the figure, the dilaton charge is monotonic and the black hole with a larger mass has a larger scalar hair. So it is expected that by increasing $\alpha$ in EMD black holes the position of ISCO radius is shifted towards the larger radii which is in agreement with table 1.

\begin{figure}[H]
\centering
\includegraphics[width=3.0in]{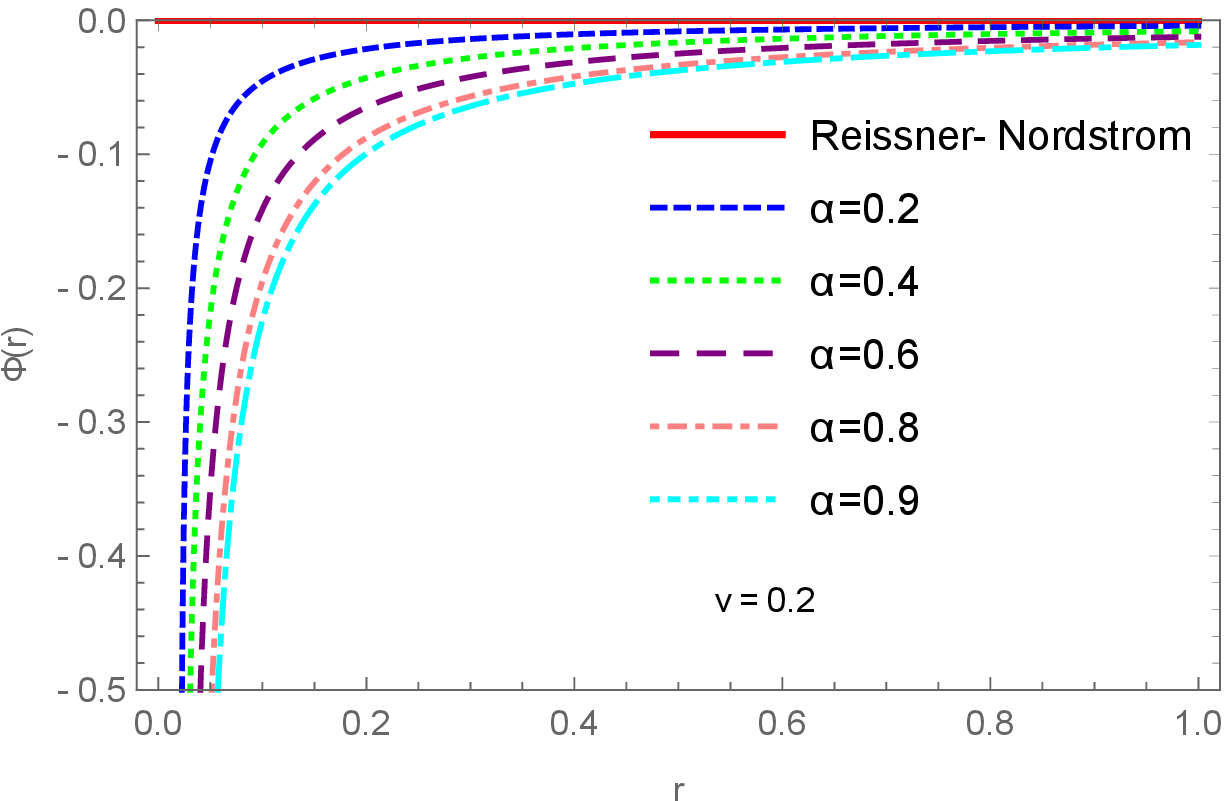}
\includegraphics[width=3.0in]{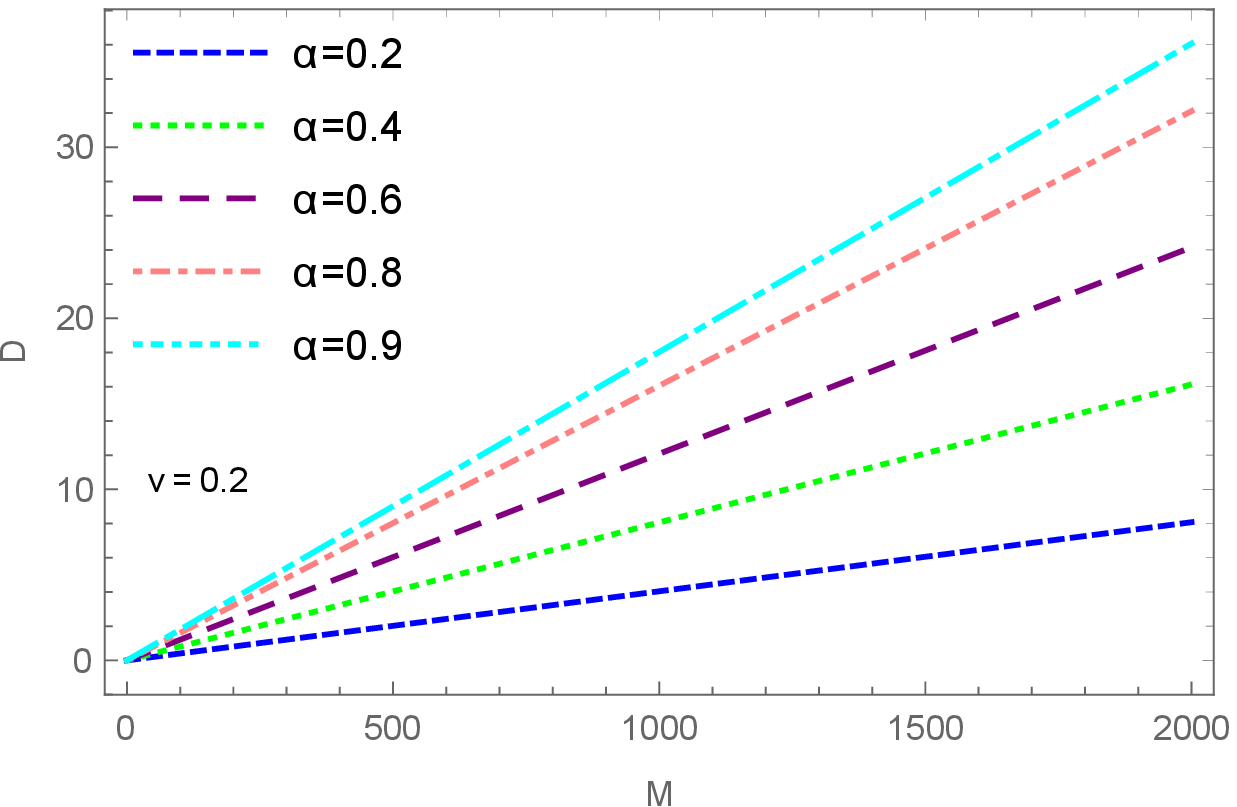}
\caption{The amplitude of the scalar field $\Phi(r)$, left panel, and dilaton charge $D$, right panel, are plotted against values of dilaton parameter and are compared to a Reissner-Nordstrom black hole ($\alpha=0$). Parameter $v$ is set to $v=0.2$.}
\label{scalar}
\end{figure}

\subsection{Charged rotating dilaton black holes}
Let us now study the physical properties of accretion  disks around charged rotating dilaton black holes in EMDG. Here, there are only two exact black hole solutions for particular values of dilaton parameter; Kerr-Newmann ($\alpha=0$) and Kaluza-Klein ($\alpha=\sqrt{3}$) \cite{Kaluza}. However, for arbitrary values of coupling constant $\alpha$, an approximate solution for slowly rotating black holes has been constructed in \cite{d22}. It is found that perturbing the rotation parameter $a$ by an infinitesimal value to order ${\cal O}(a)$ only causes components of the metric and gauge field to change and thus behavior of the dilaton field dose not change to first order. Therefore, for an infinitesimal rotation parameter, the metric of a charged rotating dilaton black hole is given by \cite{d22}
\begin{equation}
ds^2=-f(r)dt^2+\frac{dr^2}{f(r)}-2aU(r)\sin^2\theta dtd\phi+R^2(r)\left(d\theta^2+\sin ^2\theta d\phi^2\right),
\label{28}
\end{equation}
with
\begin{equation}
U(r)=\frac{r^2(1+\alpha^2)^2(1-\frac{r_-}{r})^{\frac{2\alpha^2}{1+\alpha^2}}}{r_-^2(1-\alpha^2)(1-3\alpha^2)}-
\left(1-\frac{r_-}{r}\right)^{\frac{1-\alpha^2}{1+\alpha^2}}
\left(1+\frac{r^2(1+\alpha^2)^2}{r_-^2(1-\alpha^2)(1-3\alpha^2)}+\frac{r(1+\alpha^2)}{r_{-}(1-\alpha^2)}-
\frac{r_{+}}{r}\right).\label{29}
\end{equation}
Although $U(r)$ appears to be singular at $\alpha=1$ and $\alpha=\frac{1}{\sqrt{3}}$, it is well behaved and approaches a finite value in each case. Similar to static dilaton black holes, the extreme limit of solution (\ref{28}) is given by $Q_{max}=M\sqrt{1+\alpha^2}$. Also in \cite{super}--\cite{Maeda} it is shown that there is a critical value of $\alpha\sim1$ for which  black holes may evolve into having a naked singularity for $\alpha>1$. The vector potential has the following form
\begin{equation}
A_{\phi} =-a\sin^2\theta \frac{Mv}{r}.
\label{30}
\end{equation}
One notes that for $\alpha=0$, the above solution reduces to the well-known Kerr-Newmann solution. Also, for any $\alpha$, an uncharged ($v=0$) rotating black hole reduces to Kerr solution.

At this point it is appropriate to investigate how different parameters influence the physical properties of thin disks around slowly rotating, charged dilaton black holes. In figure 5, we have plotted the flux distribution for these black holes for given values of $\alpha$ and $v$. The rotation parameter runs from 0.1 to 0.3 in order to see its effects on the energy flux of thin disks. The plot shows that as the rotation parameter increases, the energy flux also increases.

Similarly, the energy flux and temperature distribution for a fixed value of the rotation parameter, $a=0.1$, are shown in figure 6 and compared to Kerr black hole in GR. As expected, the energy flux is always larger for a charged, slowly rotating black hole in EMDG than the Kerr black hole in GR and with increasing $v$, the maximum value of the energy flux also increases so that close to extremity EMD black holes have the largest maximum of the energy flux.

\begin{figure}[H]
\centering
\includegraphics[width=3.0in]{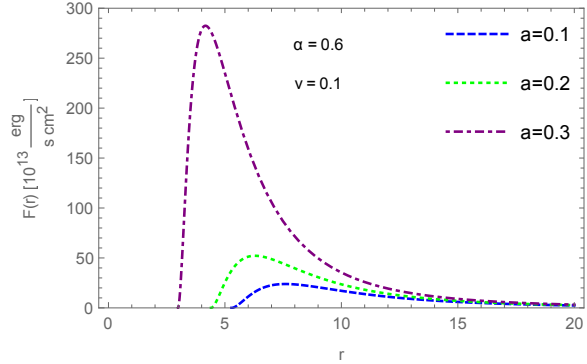}\\
\caption{The energy flux $F(r)$ from a disk around a charged, slowly rotating dilaton black hole with mass accretion rate $\dot{M}=2\times10^{-6}M_{\odot}yr^{-1}$, for $\alpha=0.6$, $v=0.1$ and different values of rotation parameter $a=0.1, 0.2, 0.3$.}
\label{fluxx}
\end{figure}

\begin{figure}[H]
\centering
\includegraphics[width=3.0in]{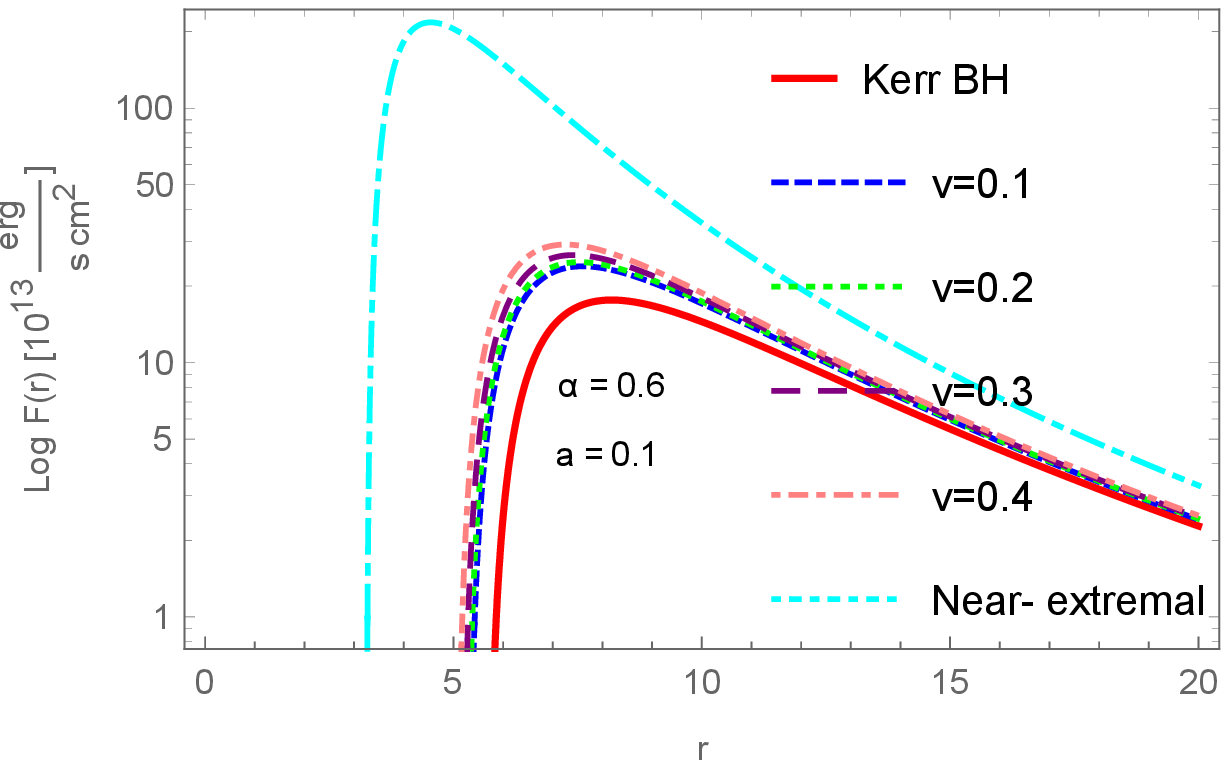}
\includegraphics[width=3.0in]{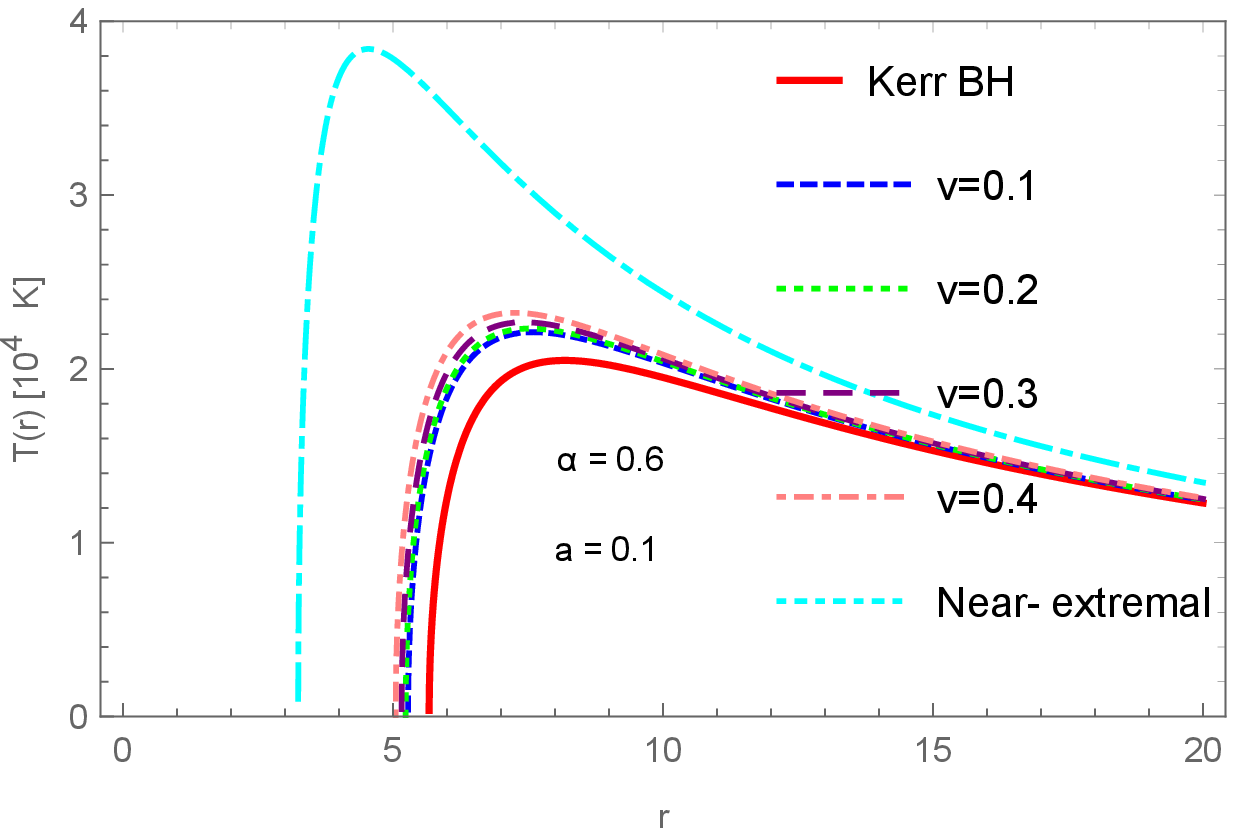}\\
\caption{The energy flux $F(r)$, left panel, and disk temperature $T(r)$, right panel, for a charged, slowly rotating dilaton black hole with mass accretion rate $\dot{M}=2\times10^{-6}M_{\odot}yr^{-1}$, for different values of $v$. In each panel the solid curve correspond to Kerr ($v=0$) black holes. At the near-extremal limit, $Q_{max}\sim M\sqrt{1+\alpha^2}$, the energy flux and disk temperature have the largest maxima. The dilaton parameter is set to $\alpha=0.6$ and rotation parameter to $a=0.1$.}
\label{temp}
\end{figure}

In table 2, we also present the ISCO radius and efficiency of a charged, slowly rotating dilaton  black hole for different values of $a$ and $v$. For a fixed value of $v$,  the ISCO radius decreases as the value of rotation parameter increases. Therefore, as one expects, the amount of energy radiated away by the accretion disk increases which is in agreement with figure 5. We also see that for a given value of the rotation parameter, the ISCO radius increase as the value of $v$ decreases, so that for a Kerr black hole with smallest value of charge ($v=0$), the ISCO radius has the largest value, in contrast to energy flux, temperature and efficiency for which it has the lowest values, as shown in figure 6. This result is due to the fact that the gravitational field in GR is stronger than that in EMDG. Therefore, the instability area around the black hole increases and the ISCO radius takes larger values. Similar to a static dilaton black hole, the results for an extremal black hole in table 3 shows that close to extremity, slowly rotating dilaton black holes are more efficient as  shown in figure 6.

Finally, it is worth stressing that the effect of dilaton parameter $\alpha$ on the disk properties is the same as that in previous section and we therefore set $\alpha=0.6$ in table 2.

\begin{table}[H]
\centering
\caption{The $r_{isco}$ of the accretion disk and efficiency for charged, slowly rotating dilaton black holes.}
\begin{tabular}{|l|l|l|l|l|}
\hline
$a$&$v$&$r_{isco}/M$& $r_{+}/M$&$\epsilon$\\ [0.5ex]
\hline
{0.1}
     &0& 5.6693&1.9945&0.0606\\
     &0.1& 5.2789&1.9968&0.0652\\
     &0.2& 5.2342&1.9871&0.0658\\
     &0.3& 5.1585&1.9708&0.0668\\
     &0.4& 5.0499&1.9474&0.0683\\
\hline
{0.2}
&0& 5.3294&1.9786&0.0646\\
&0.1& 4.4017&1.9968&0.0785\\
&0.2& 4.3472&1.9871&0.0796\\
&0.3&4.2539&1.9708&0.0814\\
&0.4& 4.1179&1.9474&0.0842\\
\hline
{0.3}
&0& 4.9786&1.9539&0.0694\\
&0.1& 2.9840&1.9968&0.1153\\
&0.2&2.8729&1.9871&0.1197\\
&0.3&2.6574&1.9708&0.1291\\
&0.4&2.1972&1.9474&0.1537\\
\hline
\end{tabular}
\end{table}

\begin{table}[H]
\centering
\caption{The $r_{isco}$ of the accretion disk and efficiency for static, charged, slowly rotating dilaton black holes at the extremal limit.}
\begin{tabular}{|l|l|l|l|l|l|}
\hline
$a$&$\alpha$&$Q_{max}/M$&$r_{isco}/M$& $r_{+}/M$&$\epsilon$\\ [0.5ex]
\hline
{0}
     &0.2&1.0198&3.9464&1.04&0.0832\\
     &0.4&1.0770&3.7825&1.16&0.0889\\
     &0.6&1.1662&3.4958&1.36&0.1009\\
     &0.8&1.2806&3.0474&1.64&0.1271\\
\hline
{0.1}
&0.2&1.0198&1.0737&1.04&0.7764\\
&0.4&1.0770&1.2091&1.16&0.7465\\
&0.6&1.16622&1.4173&1.36&0.7334\\
&0.8&1.2806&1.7012&1.64&0.7147\\

\hline
\end{tabular}
\end{table}

\section{Conclusions}
In this paper, we have studied the accretion process of thin disks around static charged dilaton black holes obtained by Gibbons and Maeda \cite{2} and also slowly rotating, charged, dilaton black holes derived by Shiraishi \cite{d22}. We applied the steady-state Novikov-Thorne model to these dilaton black holes and numerically obtained the accretion disk profiles such as the energy flux and temperature distribution for thin accretion disks. We calculated the ISCO radius for both the static and rotating EMD black holes and found that for EMD black holes the ISCO radius is smaller than the Schwarzschild and Kerr black holes. By using the ISCO radius of EMD black holes, we  also obtained the conversion efficiency of the accreting mass to radiation and shown that the EMD black holes are much more efficient  than Schwarzschild and Kerr black holes. We compared the results to that in GR and found that the accretion disks for EMDG  black holes are hotter and more luminous than in GR. As observations of electromagnetic spectrum become more accurate,  study of emission spectra of accretion disks can be a way for testing EMDG.

\end{document}